\documentclass[twocolumn,showpacs]{revtex4}
\usepackage{graphicx}

\begin{document}
\title{Angular momentum conservation for uniformly expanding flows}
\author{Sean A. Hayward}
\affiliation{Center for Astrophysics, Shanghai Normal University, 100 Guilin
Road, Shanghai 200234, China}
\date{3rd November 2006}

\begin{abstract}
Angular momentum has recently been defined  as a surface integral
involving an axial vector and a twist 1-form, which measures the
twisting around of space-time due to a rotating mass. The axial
vector is chosen to be a transverse, divergence-free, coordinate
vector, which is compatible with any initial choice of axis and
integral curves. Then a conservation equation expresses rate of
change of angular momentum along a uniformly expanding flow as a
surface integral of angular momentum densities, with the same form
as the standard equation for an axial Killing vector, apart from the
inclusion of an effective energy tensor for gravitational radiation.
\end{abstract}
 \maketitle

\section{Introduction}
Energy, momentum and angular momentum have long been issues in
General Relativity, since it is difficult to find definitions with
satisfactory properties for a general space-time, though they make
sense for weak gravitational fields, with physical properties
familiar from flat space-time \cite{MTW}. For an asymptotically flat
space-time, there is an accepted notion of total energy-momentum at
infinity, satisfying the Bondi energy-momentum flux equation,
whereas angular momentum is more delicate: there is an accepted
definition at spatial infinity but not at null infinity, and
therefore no accepted flux equation. In the strong-field regime,
flux equations in the form of conservation laws have recently been
found for black holes, specifically for trapping horizons
\cite{bhd}, for both energy \cite{bhd2,bhd3} and angular momentum
\cite{bhd4,bhd5}.

All these results measure energy or angular momentum on a spatial
surface, usually of spherical topology, with flux equations
concerning a flow of such surfaces. For a general flow of general
surfaces in a general space-time, it seems unlikely that physically
meaningful conservation laws exist. However, there is a
geometrically preferred type of flow which has been called uniformly
expanding \cite{mon}, for which an energy conservation law was
recently found \cite{gr,BHMS}, with the same form as that for black
holes. This article derives the corresponding conservation law for
angular momentum, which also has the same form as that for black
holes.

The article is organized as follows. Sections II, III and IV
respectively review uniformly expanding flows, a dual-null formalism
and conservation of energy. Section V describes angular momentum,
Section VI its conservation and Section VII concludes.

\section{Uniformly expanding flows}
\begin{figure}
\includegraphics[height=15mm]{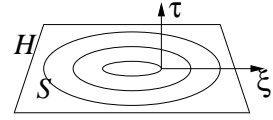}
\caption{A non-null hypersurface $H$ foliated by spatial surfaces
$S$, with generating vector $\xi$ and its normal dual $\tau=\xi^*$.
If $H$ becomes null, $\xi$ and $\tau$ coincide.} \label{normal}
\end{figure}

General Relativity will be assumed, with space-time metric $g$.
Consider a flow of spatial surfaces $S$, i.e.\ a one-parameter
family $\{S\}$, locally generating a foliated hypersurface $H$.
Labelling the surfaces by a coordinate $x$, they are generated by a
flow vector $\xi=\partial/\partial x$, which can be taken to be
normal to the surfaces, $\bot\xi=0$, where $\bot$ denotes projection
onto $S$. Given a normal vector $v$, $\bot v=0$, a Hodge duality
operation yields a dual normal vector $v^*$ satisfying
\begin{equation}
\bot v^*=0,\quad g(v^*,v)=0,\quad g(v^*,v^*)=-g(v,v).\label{nd}
\end{equation}
In particular,
\begin{equation}
\tau=\xi^*
\end{equation}
is normal to $H$, with the same scaling as $\xi$ (Fig.\ref{normal}). The
coordinate freedom here is just $x\mapsto\tilde x(x)$ and choice of transverse
coordinates on $S$, under which all the key formulas will be invariant.

The expansion 1-form
\begin{equation}
\theta={*}d{*}1\label{exp}
\end{equation}
where $*$ denotes the Hodge operator of $S$ and $d$ the exterior derivative in
the normal space, yields the expansion
\begin{equation}
\theta_v=\theta(v)={*}L_v{*}1
\end{equation}
along a normal vector $v$, where $L$ denotes the Lie derivative. One can also
introduce the unit-expansion vector
\begin{equation}
\eta=g^{-1}(\theta)/g^{-1}(\theta,\theta)
\end{equation}
where the metric sign convention is that spatial metrics are positive
definite. Then $\eta$ and its dual $\eta^*$ have expansions
\begin{equation}
\theta_\eta=1,\quad\theta_{\eta^*}=0.
\end{equation}
Other names for $g^{-1}(\theta)$ and $\eta$ are mean-curvature vector and
inverse mean-curvature vector respectively.

A {\em uniformly expanding flow} \cite{mon,gr,BHMS} is a one-parameter family
of surfaces on which both the expansion $\theta_\xi$ of the flow and the dual
expansion $\theta_\tau$ are constant on the surfaces:
\begin{equation}
D\theta_\xi=D\theta_\tau=0.\label{uef}
\end{equation}
Equivalently, such flows are given by
\begin{equation}
\xi=a(\eta+c\eta^*),\quad Da=Dc=0
\end{equation}
and thereby generalize inverse mean-curvature flows from codimension 1 to
codimension 2.

\section{Dual-null formalism}
It is convenient to use a dual-null formalism \cite{dn,dne},
describing two families of null hypersurfaces $\Sigma_\pm$,
intersecting in a two-parameter family of spatial surfaces,
including the desired one-parameter family. Relevant aspects of the
formalism are summarized as follows. This material is not new, but
included to provide a self-consistent treatment.

Labelling $\Sigma_\mp$ by coordinates $x^\pm$ which increase to the
future, one may take transverse coordinates $x^a$ on $S$, which for
a sphere would normally be angular coordinates
$x^a=(\vartheta,\varphi)$. Writing space-time coordinates
$x^\alpha=(x^+,x^-,x^a)$ indicates how one may use Greek letters
$(\alpha,\beta,\ldots)$ for space-time indices and corresponding
Latin letters $(a,b,\ldots)$ for transverse indices. The coordinate
basis vectors are $\partial_\alpha=\partial/\partial x^\alpha$ and
the dual 1-forms are $dx^\alpha$, satisfying
$\partial_\beta(dx^\alpha)=\delta^\alpha_\beta$. Coordinate vectors
commute, $[\partial_\alpha,\partial_\beta]=0$, where the brackets
denote the Lie bracket or commutator. Two coordinate vectors have a
special role, the evolution vectors $\partial_\pm=\partial/\partial
x^\pm$ which generate the dynamics, spanning an integrable evolution
space. The corresponding normal 1-forms $dx^\pm$ are null by
assumption:
\begin{equation}
g^{-1}(dx^\pm,dx^\pm)=0.
\end{equation}
The relative normalization of the null normals may be encoded in a function
$f$ defined by
\begin{equation}
e^f=-g^{-1}(dx^+,dx^-).\label{ff}
\end{equation}
The transverse metric, or the induced metric on $S$, is found to be
\begin{equation}
h=g+2e^{-f}dx^+\otimes dx^-\label{tm}
\end{equation}
where $\otimes$ denotes the symmetric tensor product. There are two shift
vectors
\begin{equation}
s_\pm=\bot\partial_\pm
\end{equation}
where $\bot$ is extended to denote projection by $h$. The null normal vectors
\begin{equation}
l_\pm=\partial_\pm-s_\pm=-e^{-f}g^{-1}(dx^\mp)\label{nn}
\end{equation}
are future-null and satisfy
\begin{eqnarray}
&&g(l_\pm,l_\pm)=0,\quad g(l_+,l_-)=-e^{-f},\nonumber\\
&&l_\pm(dx^\pm)=1,\quad l_\pm(dx^\mp)=0,\quad\bot l_\pm=0.\label{ls}
\end{eqnarray}
The metric takes the form
\begin{eqnarray}
g&=&h_{ab}(dx^a+s_+^adx^++s_-^adx^-)\otimes\nonumber\\&&(dx^b+s_+^bdx^++s_-^bdx^-)
-2e^{-f}dx^+\otimes dx^-.
\end{eqnarray}
Then $(h,f,s_\pm)$ are configuration
fields and the independent momentum fields are found to be linear combinations
of the following transverse tensors:
\begin{eqnarray}
\theta_\pm&=&{*}L_\pm{*}1\label{ex}\\ \sigma_\pm&=&\bot L_\pm h-\theta_\pm h\label{sh}\\
\nu_\pm&=&L_\pm f\label{in}\\
\omega&=&{\textstyle\frac12}e^fh([l_-,l_+])\label{tw}
\end{eqnarray}
where $L_\pm$ is shorthand for the Lie derivative along $l_\pm$. Adding
indices explicitly, the functions $\theta_\pm$ are the null expansions, the
traceless bilinear forms $\sigma_{\pm ab}$ are the null shears, the 1-form
$\omega_a$ is the twist, measuring the lack of integrability of the normal
space, and the functions $\nu_\pm$ are the inaffinities, measuring the failure
of the null normals to be affine. The fields
$(\theta_\pm,\sigma_\pm,\nu_\pm,\omega)$ encode the extrinsic curvature of the
dual-null foliation. These extrinsic fields are unique up to interchange
$\pm\mapsto\mp$ and diffeomorphisms $x^\pm\mapsto\tilde x^\pm(x^\pm)$ which
relabel the null hypersurfaces. It will also be convenient to use capital
Latin letters $(A,B,\ldots)$ for normal indices, when not denoted by $\pm$ in
the dual-null basis. Then the configuration fields are $(h_{ab},f,s_A{}^b)$,
the momentum fields are $(\theta_A,\sigma_{Abc},\nu_A,\omega_a)$ and the
derivative operators are $(\bot L_A,D_a)$, where $D$ is the covariant
derivative operator of $h$.

Returning to a general foliated hypersurface $H$, a normal vector $v$ has
components $v^\pm$ along $l_\pm$, so that $v=v^+l_++v^-l_-$, and its normal
dual is $v^*=v^+l_+-v^-l_-$. In particular, the generating vector is
\begin{equation}
\xi=\xi^+l_++\xi^-l_-\label{xi}
\end{equation}
and its dual is
\begin{equation}
\tau=\xi^+l_+-\xi^-l_-.\label{tau}
\end{equation}
Since $H$ is given parametrically by functions $x^\pm(x)$, the
components $\xi^\pm=\partial x^\pm/\partial x$ are constant on the
surfaces:
\begin{equation}
D\xi^\pm=0.\label{dxi}
\end{equation}
The expansion 1-form (\ref{exp}) is given by
\begin{equation}
\theta=\theta_+dx^++\theta_-dx^-
\end{equation}
so that the expansion along a normal vector $v$ can be expressed as
\begin{equation}
\theta_v=\theta_Av^A
\end{equation}
and in particular, $\theta_\xi=\xi^+\theta_++\xi^-\theta_-$ and
$\theta_\tau=\xi^+\theta_+-\xi^-\theta_-$. The conditions (\ref{uef}) defining
a uniformly expanding flow can then be expressed in terms of the null
expansions as
\begin{equation}
D\theta_+=D\theta_-=0
\end{equation}
unless the flow is null, in which case $\xi=\pm\tau$ and the two expansion
conditions become one.

\section{Conservation of energy}
The energy conservation law \cite{mon,gr,BHMS} will be stated here for later
comparison, modifying some notation. Assuming compact $S$ henceforth, the
transverse surfaces have area
\begin{equation}
A=\oint_S{*}1
\end{equation}
and the area radius
\begin{equation}
R=\sqrt{A/4\pi}
\end{equation}
is often more convenient. The Hawking mass \cite{Haw}
\begin{equation}
M=\frac
R2\left(1-\frac1{16\pi}\oint_S{*}g^{AB}\theta_A\theta_B\right)\label{hm}
\end{equation}
can be used as a measure of the active gravitational mass or energy on a
transverse surface. Here units are such that Newton's gravitational constant
is unity.

An energy conservation equation requires a vector playing the role of a
stationary Killing vector. For a general compact surface, the simplest
definition of such a vector which becomes unit for round spheres in flat
space-time is \cite{bhd2,bhd3}
\begin{equation}
k=(g^{-1}(dR))^*.\label{time}
\end{equation}
This vector actually was found to be the appropriate dual of $M$, in the sense
of conservation laws for black holes \cite{bhd2,bhd3} and for uniformly
expanding flows \cite{gr,BHMS}. In either case, the energy conservation law
can be written as
\begin{equation}
L_\xi M\cong\oint_S{*}(T_{AB}+\Theta_{AB})k^A\tau^B \label{ec}
\end{equation}
where $\Theta$ is an effective energy tensor for gravitational radiation. This
determines only the normal-normal components of $\Theta$, as
\begin{eqnarray}
\Theta_{\pm\pm}&=&||\sigma_\pm||^2/32\pi\label{Theta0}\\
\Theta_{\pm\mp}&=&e^{-f}|\omega\mp{\textstyle\frac12}Df|^2/8\pi\label{Theta1}
\end{eqnarray}
where $|\zeta|^2=h^{ab}\zeta_a\zeta_b$ and
$||\sigma||^2=h^{ac}h^{bd}\sigma_{ab}\sigma_{cd}$. Further discussion is
referred to \cite{bhd2,bhd3,gr,BHMS}.

\begin{figure}
\includegraphics[height=3cm]{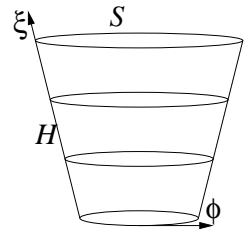}
\caption{A transverse vector $\psi$.} \label{transverse}
\end{figure}

\section{Angular momentum}
The angular momentum of a surface in a flow, as a functional of an
axial vector $\psi$, is defined following \cite{bhd4,bhd5} as
\begin{equation}
J[\psi]=\frac1{8\pi}\oint_S{*}\psi^a\omega_a \label{am}
\end{equation}
where $\omega$ is the twist (\ref{tw}). Since the twist encodes the
non-integrability of the normal space, it provides a geometrical
measure of the twisting around of space-time due to rotational
frame-dragging. It is an invariant of a dual-null foliation and
therefore of a non-null foliated hypersurface $H$, so $J[\psi]$ is
also an invariant. The definition was obtained from the Komar
integral \cite{Kom} and shown to recover the standard definition of
angular momentum for a weak-field metric \cite{MTW}, with the twist
being directly related to the precessional angular velocity of a
gyroscope due to the Lense-Thirring effect.

Following \cite{bhd4,bhd5}, the axial vector will be assumed to be transverse,
$\bot\psi=\psi$ (Fig.\ref{transverse}), to be a coordinate vector
$\psi=\partial/\partial\varphi$ on $H$, implying
\begin{equation}
L_\xi\psi\cong0 \label{lie}
\end{equation}
and to have vanishing transverse divergence:
\begin{equation}
D_a\psi^a\cong0. \label{div}
\end{equation}
The last condition holds if $\psi$ is an axial Killing vector, and can be
understood as a weaker condition, equivalent to $\psi$ generating a symmetry
of the area form rather than of the whole metric, since
$L_\psi({*}1)={*}D_a\psi^a$. Alternatively, assuming that the integral curves
$\gamma$ of $\psi$ are closed, it can always be satisfied by choice of scaling
of $\psi$. It implies that $\psi$ is locally the curl of some function, whose
equipotentials are $\gamma$.

\begin{figure}
\includegraphics[height=25mm]{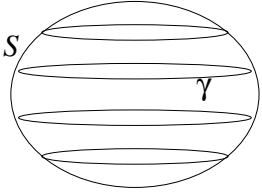}
\caption{Integral curves $\gamma$ of an axial vector $\psi$.} \label{axial}
\end{figure}

In the situation normally envisaged for angular momentum, $S$ would have
spherical topology and $\gamma$ would form a smooth foliation of circles,
covering the surface except for two poles (Fig.\ref{axial}), with the
coordinate $\varphi$ identified at 0 and $2\pi$. In fact, such conditions play
no role in the conservation law to be derived in the next section, which
requires only the conditions (\ref{lie})--(\ref{div}) on transverse $\psi$.
Note also that the commutator identity \cite{dne}
\begin{equation}
L_\xi(D_a\psi^a)-D_a(L_\xi\psi)^a=\psi^aD_a\theta_\xi \label{com}
\end{equation}
which then forces
\begin{equation}
\psi^aD_a\theta_\xi\cong0 \label{con}
\end{equation}
is automatically satisfied for uniformly expanding flows (\ref{uef}). This can
be seen as a motivation for the divergence-free condition (\ref{div}), since
(\ref{lie}) already forces $L_\xi(D_a\psi^a)\cong0$ for a uniformly expanding
flow. This is quite different to the situation for black holes
\cite{bhd4,bhd5}. Thus there is a large family of conservation laws,
corresponding to angular momentum about different axes. In the case of
spherical topology, one can choose any two points on an initial surface as
poles, and any set of circles $\gamma$ interpolating smoothly between them,
then there will exist a vector $\psi$ with integral curves $\gamma$ satisfying
(\ref{div}). The condition (\ref{lie}) then propagates $\psi$ along the flow.

\section{Conservation of angular momentum}
The rate of change of angular momentum (\ref{am}) along the flow, using
(\ref{lie}), is
\begin{equation}
L_\xi
J\cong\frac1{8\pi}\oint_S{*}(\theta_\xi\psi^a\omega_a+\psi^aL_\xi\omega_a).
\end{equation}
The twisting equations
\begin{eqnarray}
\bot L_\pm\omega_a&=&-\theta_\pm\omega_a\pm\textstyle{\frac12}D_a\nu_\pm
\mp\textstyle{\frac12}D_a\theta_\pm\mp\textstyle{\frac12}\theta_\pm
D_af \nonumber\\
&&\qquad\qquad{}\pm\textstyle{\frac12}h^{cd}D_d\sigma_{\pm ac} \mp8\pi
T_{a\pm}\label{twisting}
\end{eqnarray}
were obtained from the Einstein equations, where
$T_{a\pm}=h_a^\gamma T_{\gamma\beta}l_\pm^\beta$ is the
transverse-normal projection of the energy tensor $T$ \cite{gwbh}.
They can be used to express
\begin{eqnarray}
\bot L_\xi\omega_a&=&-\theta_\xi\omega_a+\tau^B(\textstyle{\frac12}D_a\nu_B
-\textstyle{\frac12}D_a\theta_B-\textstyle{\frac12}\theta_BD_af \nonumber\\
&&\qquad\qquad\quad{}+\textstyle{\frac12}h^{cd}D_d\sigma_{Bac}-8\pi
T_{aB}).
\end{eqnarray}
Then
\begin{eqnarray}
L_\xi J&\cong&-\frac1{8\pi}\oint_S{*}\psi^a\tau^B\Big(8\pi
T_{aB}-\textstyle{\frac12}h^{cd}D_d\sigma_{Bac}\nonumber\\
&&\qquad\quad{}-\textstyle{\frac12}\left(D_a\nu_B
-D_a\theta_B-\theta_BD_af\right)\Big).
\end{eqnarray}
Now the terms in $D\nu$ and $D\theta$ may be removed as total divergences due
to (\ref{dxi}) and (\ref{div}). To remove the term in $\theta Df$ generally
requires
\begin{equation}
\psi^aD_a\theta_\tau\cong0.
\end{equation}
Clearly this would be generally inconsistent with (\ref{con}), specifically
if $\theta_\xi$ and $\theta_\tau$ have different equipotentials. However, it
is consistent in three special cases: (i) for null $\xi$, since then
$\xi=\pm\tau$; (ii) along a trapping horizon $\theta_\pm\cong0$, since then
$\theta_\xi=\mp\theta_\tau$ \cite{bhd4,bhd5}; and (iii) for a uniformly
expanding flow (\ref{uef}). In each case, this leaves just the $T$ and
$\sigma$ terms, with the latter expressible in terms of the transverse-normal
block
\begin{equation}
\Theta_{aB}=-\frac1{16\pi}h^{cd}D_d\sigma_{Bac} \label{Theta2}
\end{equation}
of the effective energy tensor for gravitational radiation. Finally the
desired conservation law for angular momentum is obtained as
\begin{equation}
L_\xi J\cong-\oint_S{*}(T_{aB}+\Theta_{aB})\psi^a\tau^B. \label{amc}
\end{equation}
Apart from the inclusion of $\Theta$, this is the standard surface-integral
form of conservation of angular momentum, were $\psi$ an axial Killing vector.
The null shears $\sigma_{\pm bc}$ have previously been identified in the
energy conservation law (\ref{ec}) as encoding the ingoing and outgoing
transverse gravitational radiation, via the energy densities (\ref{Theta0})
\cite{bhd2,bhd3}. So the expression (\ref{Theta2}) implies that gravitational
radiation with a transversely differential waveform will generally possess
angular momentum density.

\section{Conclusion}

The conservation laws (\ref{ec}) and (\ref{amc}) take a similar
form, expressing rate of change of mass $M$ and angular momentum $J$
as surface integrals of densities of energy and angular momentum,
with respect to vectors $k$ and $\psi$ which play the role of
stationary and axial Killing vectors. They also take the same form
as the corresponding conservation laws for black holes
\cite{bhd2,bhd3,bhd4,bhd5}, with the effective energy tensor
$\Theta$ for gravitational radiation also taking the same form
(\ref{Theta0}), (\ref{Theta1}), (\ref{Theta2}).

This provides further evidence for the utility of the definition
(\ref{am}) of angular momentum in terms of the twist, and of
uniformly expanding flows. However, the question of existence of
such flows is generally unresolved \cite{BHMS} except for a
time-symmetric hypersurface, where weak flows have been proven to
exist \cite{Bra,HI}. Another caveat is that an initial surface
should be chosen carefully, in particular not being highly
distorted, if the physical interpretation is to be plausible. While
no comprehensive prescription is known, it would be reasonable to
restrict to topologically spherical surfaces with positive Gaussian
curvature. Given a black hole, one could choose a marginal surface
to be propagated outward. In an asymptotically flat space-time, one
could could propagate inward a suitable surface at null infinity,
such as those provided by the recent unified framework for null and
spatial infinity \cite{inf}. Since the Bondi energy flux law can
also be written in the conservation form (\ref{ec}), this also
offers hope to resolve the issue of angular momentum at null
infinity.

Research supported by the National Natural Science Foundation of
China under grants 10375081 and 10473007, and by Shanghai Normal
University under grant PL609.

\appendix

\end{document}